# Dynamic wake modulation induced by utility-scale wind turbine operation

Aliza Abraham[1,2] and Jiarong Hong[1,2]

[1]St. Anthony Falls Laboratory, College of Science and Engineering, University of Minnesota, Minneapolis, MN, USA.
[2]Department of Mechanical Engineering, College of Science and Engineering, University of Minnesota, Minneapolis, MN, USA.

**Abstract**

Understanding wind turbine wake mixing and recovery is critical for improving the power generation and structural stability of downwind turbines in a wind farm. In the field, where incoming flow and turbine operation are constantly changing, wake recovery can be significantly influenced by dynamic wake modulation, which has not yet been explored. Here we present the first investigation of dynamic wake modulation in the near wake of a utility-scale turbine and quantify its relationship with changing conditions. This investigation is enabled using novel super-large-scale flow visualization with natural snowfall, providing unprecedented spatiotemporal resolution to resolve instantaneous changes of the wake envelope. These measurements reveal the significant influence of dynamic wake modulation on wake recovery. Further, our study uncovers the direct connection of dynamic wake modulation with operational parameters readily available to the turbine, paving the way for more precise wake prediction and control under field conditions for wind farm optimization.

Wind turbine wakes, the regions of slower and more turbulent air behind turbines, can lead to significant power loss within and even between adjacent wind farms (Barthelmie et al. 2009, Lundquist et al. 2019). Therefore, an improved understanding of the utility-scale wind turbine wake and its development is critical for the optimization of wind farm layout and controls. One of the most important aspects of wake development is wake recovery, i.e., the return of the region downwind of the turbine to ambient flow conditions, caused by re-entrainment of momentum through mixing with the surrounding flow. This process is critical for increasing the amount of kinetic energy available for downwind power generation (Calaf et al. 2010, Cal et al. 2010, Peña & Rathmann 2014). It has been well recognized that several factors, including the breakdown of vortex structures shed from the blades (Lignarolo et al. 2015), the size and spacing of the turbines (Vermeer et al. 2003), and the stability of the atmosphere (Magnusson & Smedman 1994, Hansen et al. 2012), influence wake recovery. However, dynamic wake modulation, referring to large-scale dynamic motion of the wake, including instantaneous wake expansion and deflection, can also significantly affect its recovery but has not yet been explored. This phenomenon is typically induced by constantly changing turbine operation and incoming flow conditions associated with utility-scale turbines in the field. Specifically, the amount of wake expansion is determined by axial induction, i.e., the change in streamwise velocity when passing through the rotor, which is in turn a function of the turbine thrust coefficient. Therefore, the changes in axial induction due to variations in incoming flow and turbine operation can induce instantaneous fluctuation in the wake expansion angle. In the extreme case, wake contraction can occur when the turbine undergoes changes in blade pitch (Dasari et al. 2019), significantly affecting the extent of the wake and the rate of wake recovery. Wake deflection occurs when a turbine is operating with yaw error, i.e., the rotor direction is not aligned with the wind direction, and a spanwise force (parallel to the ground and perpendicular to the wind direction) is exerted on the wake, causing it to deflect in the spanwise direction (Jiménez et al. 2010, Fleming et al. 2017). Similar to wake expansion, the deflection angle of the wake can also fluctuate in response to changing wind direction, which enhances mixing and modifies the characteristics of wake recovery.

Although a significant number of wake studies (Jiménez et al. 2010, Chamorro & Porté-Agel 2010, Smalikho et al. 2013, Aitken et al. 2014, Fleming et al. 2014, Mirocha et al. 2015, Annoni et al. 2016, Machefaux et al. 2016, Bromm et al. 2018) have been conducted to understand the wake behavior under different wind conditions, very few have accounted for the effect of dynamic wake modulation on wake development. The reason is that such instantaneous changes in conditions are difficult to model on the laboratory scale and in simulations, and the associated instantaneous change of wake characteristics cannot be captured with conventional field measurement techniques (e.g., lidar, radar, sodar, etc.) due to their lack of sufficient spatial and temporal resolution. Nevertheless, some studies have investigated the effect of changing turbine operation and flow conditions on wake behavior. Specifically, continuous regulation of axial induction through blade pitch and generator torque (Marden et al. 2013, Gebraad & van Wingerden 2014) or thrust coefficient (Goit & Meyers 2015) to maximize overall wind farm power generation has been investigated. However, these studies only accounted for the changes in wake velocity, and not for changes in wake expansion and their effect on wake mixing. For wake deflection caused by changes in wind direction, the current state-of-the-art assumes the wake is released from the rotor and passively advected by large-scale turbulence (Larsen et al. 2008, Trujillo et al. 2011, Larsen et al. 2013, Vollmer et al. 2016). However, the wake is not a passive tracer; rather, the turbine actively modulates the incoming flow to influence wake behavior. Controlling wakes using yaw error alone (Ahmad et al. 2019) and a combination of both axial induction and yaw error (Munters & Meyers 2018) has also been proposed as a method for optimizing overall wind farm performance. However, to effectively implement these strategies, we need to first understand the impact on wake behavior of the constantly-changing flow and turbine operational conditions present in the field.

Consequently, here we present the first investigation of the phenomenon of dynamic wake modulation and its connection with instantaneous variation of incoming flow and turbine conditions. This study was made possible through the implementation of super-large-scale flow visualization using natural snowfall (Toloui et al. 2014, Hong et al. 2014, Nemes et al. 2017, Heisel et al. 2018, Dasari et al. 2019). This technique was first developed to characterize the atmospheric boundary layer (Toloui et al. 2014) and the near-wake of a utility-scale wind turbine (Hong et al. 2014) through imaging the motion of individual snowflakes in a field of view on the order of 10 m, providing unprecedented spatial and temporal resolution in comparison to state-of-the-art field measurement techniques. It was later extended to measure the flow field in the entire span of the turbine wake on the order of 100 m through tracking the motion of patterns of snowflakes representing vortices that are advected by the flow (Dasari et al. 2019). Here we again take advantage of the visual patterns created by vortical flow structures in snow to thoroughly characterize dynamic wake modulation by a utility-scale turbine.

**Field-scale wake visualization using snow**

The experiment was conducted at the Eolos field site in Rosemount, MN. The site consists of a 2.5 MW turbine with a 96 m rotor diameter ($D$), and a 130 m tall meteorological tower (met tower) located 170 m south of the turbine, which provides wind conditions at multiple elevations above the ground. The visualization of the near wake of the turbine was conducted at night during a snowstorm, during which the snowflakes were used as environmentally benign flow tracers, providing high spatial and temporal resolution visualization of the flow field (Supplementary Video 1). A light sheet was used to illuminate the snowflakes in a plane parallel to the rotor (perpendicular to the flow) 17 m (0.18$D$) downstream of the turbine (Fig. 1a). A high-resolution camera captured the images of snowflakes in a field of view of 73 m × 129 m (spanwise × vertical) for more than 2 hours. Coherent vortex structures in the flow appear as voids in the snow particle seeding projected onto the light sheet due to the centrifugal effect on the inertial particles caused by the strongly rotating fluid. Image enhancement techniques were used to extract the vortices, and the images were stacked to reconstruct a three-dimensional visualization showing the changes in the near-wake vortices over time (Fig. 1b). The reconstruction demonstrates the significantly different flow features below and above the hub height. Below the hub, strong interactions are observed between the vortices shed from the blades and those shed from the tower. Above the hub, the characteristic vortex helix shed from the rotating turbine blade tips is more clearly visible. Here the wake modulation becomes more evident since the tip vortex helix is not affected by the tower vortices. Therefore, an envelope was fit to the upper boundary of the portion of the blade tip vortex helix above the hub to capture the overall shape of the wake (Fig. 1c). Cross-sections of this envelope were compared to the boundary of the wake with no expansion or deflection included (referred to as the model wake hereafter), extracted from a SolidWorks model of the experimental setup. The shift between the observed wake shape and the model wake shape was decomposed into vertical and spanwise components ($\delta_{w,z}$ and $\delta_{w,y}$ respectively) at each point in time. This decomposition enabled quantification of the degree of dynamic wake modulation. The effect of these behaviors on wake propagation direction and mixing was then explored.

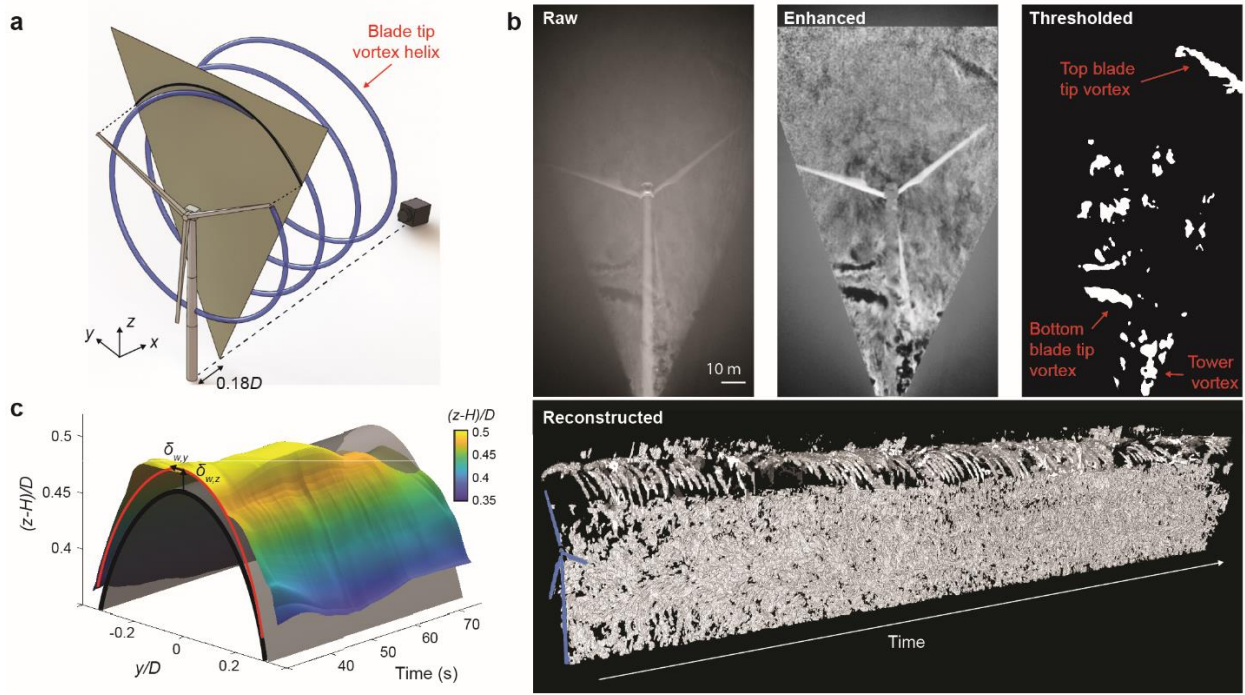

**Fig. 1 | High resolution measurement of near-wake envelope using snowflake visualization. a**, Schematic of the experimental setup including the turbine, light sheet, and camera. The blue helix represents the blade tip vortex and the black line indicates the location of the intersection between the model wake and light sheet. The coordinate system is defined, where $x$ is the streamwise direction, $y$ is the spanwise direction, and $z$ is the vertical direction. **b**, Image processing steps used to reconstruct instantaneous vortical structures in the near wake, including the raw image, enhanced image, thresholded image, and reconstructed volume. **c**, Sample time series of the wake envelope and the model wake position. The red line marks the profile of the shifted wake position least square fit to the experimental data and the arrows indicate the decomposition of the wake shift into vertical and spanwise components ($\delta_{w,z}$ and $\delta_{w,y}$, respectively).

## Intermittency of blade tip vortex formation

When examining the top tip voids in the wake reconstruction, several periods of intermittency are observed where the voids do not appear consistently (Fig. 2a). These periods are identified using the total void area within a time window. The regions with void areas corresponding to the bottom 6% within a period of four blade passes are defined as intermittent. These values were chosen because they provided the clearest separation between intermittency and consistency in the histogram including the entire data set. Tip vortex intermittency is found to correlate with turbine power and blade pitch, depending on the region of operation of the turbine. When the wind speed is low (below approximately 10 m/s), the turbine operates in region 2 or below where the control algorithm seeks to maximize the power output by keeping the blade pitch low and nearly constant, maximizing the lift on the airfoils. In this region, periods of intermittency occur when the power production is lower (Fig. 2b), because the blades generate less circulation at lower power, producing weaker tip vortices (Hong et al. 2014). Above region 2, where the pitch angle of the blades changes more, periods of intermittency occur at higher values of blade pitch, consistent with the observations of Dasari et al. (2019). This trend can be explained by the decay of tip vortex strength in response to decreasing angle of attack caused by increasing blade pitch in this region. More importantly, our results highlight the potential benefit of using information provided by the SCADA system to directly predict the occurrence of tip vortex intermittency with statistical certainty. These periods of intermittency have implications for wake recovery, as tip vortex breakdown enhances mixing (Lignarolo et al. 2015). Blade pitch and power output measurements are readily available to the turbine controller, so the knowledge of the relationship between these variables and the behavior of the tip vortices in specific regimes can be incorporated into wind farm optimization, enabling the direct modulation and prediction of wake recovery through tip vortex breakdown.

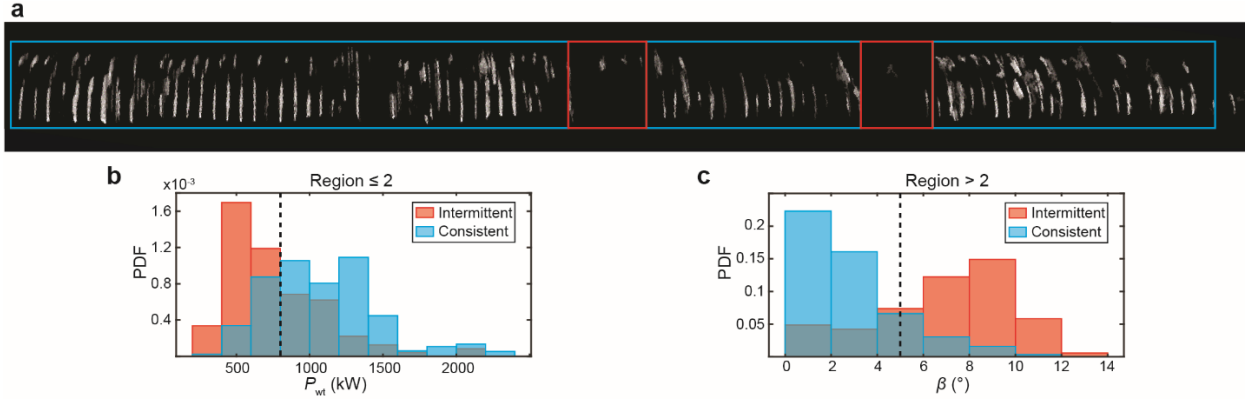

**Fig. 2 | Blade tip vortex intermittency. a**, Sample sequence of extracted top tip voids, with blue and red boxes marking periods of consistent and intermittent voids, respectively. **b**, Histogram comparing the turbine power during periods of consistent and intermittent tip vortex appearance when the turbine is operating in region 2 or below. **c**, Histogram comparing the blade pitch during periods of consistent and intermittent tip vortex appearance when the turbine is operating above region 2.

**Quantification of dynamic wake modulation**

For the analysis of dynamic wake modulation, periods of intermittency described above are removed in order to extract continuous fluctuations of the wake envelope. Figure 3 shows the spanwise wake modulation in response to instantaneous yaw error. In Fig. 3a, spanwise wake deflection is defined in terms of the displacement along the light sheet plane, $\delta_{w,y}$, and the angle of deflection, $\varphi_{w,y}$, calculated using the distance between the turbine and the light sheet. Figure 3b shows schematics comparing the wake deflection trends in response to steady yaw error and instantaneous yaw error. Under steady yaw error, the wake steering angle, $\alpha = \varphi_{w,y} - \theta$, where $\theta$ is the yaw error, is positively correlated with $\theta$ (Jiménez et al. 2010). However, under the constant wind direction changes in the field that cause instantaneous yaw error, $\alpha$ is negatively correlated with $\theta$ with a correlation coefficient of -0.42 (Fig. 3c). According to the numerical simulation from Leishman (2002), such opposite deflection can occur in the near wake of the turbine due to the fact that spanwise wake deflection takes time to stabilize under sudden changes in yaw error. Our results further suggest that, when the yaw error is short-lived as it is under the constantly changing wind conditions in the field, instantaneous wind direction changes can cause significant disturbance to the process of the wake transitioning to the fully deflected state it experiences under steady yaw error. In addition to the the inverse statistical relation compared to the steady yaw error condition, a sample time sequence also clearly shows the opposing trend of $\theta$ and $\alpha$ (Fig. 3d). More remarkably, the same inverse relationship between $\theta$ and $\alpha$ is observed (with a correlation coefficient of -0.35) when $\varphi_{w,y}$ is obtained from the spanwise velocity recorded at the met tower which, due to the northerly wind direction, is in the wake of the turbine $1.77D$ downstream. Such observation indicates the persistence further downstream of the disturbance of the spanwise wake deflection in the near wake induced by the constantly changing wind direction. To provide an assessment of the impact of spanwise dynamic wake modulation on wake position downstream, we consider the maximum steering angle caused by instantaneous yaw error, about $15°$ in this dataset. At $6D$ downstream, this steering angle corresponds to a $1.6D$ offset from the rotor axis. This shift in wake direction is significant compared to the study conducted by Vollmer et al. (2016), which found that wind direction fluctuations cause changes in wake direction up to $0.5D$ at $6D$ downstream when the wake is modelled as a passive tracer under neutrally stratified atmospheric conditions. Therefore, our results show that dynamic wake modulation has a significant effect on wake propagation direction. More importantly, we are able to demonstrate that the instantaneous yaw error, available from SCADA system of current utility-scale turbines, is strongly correlated with the spanwise wake modulation and can be used to predict wake steering and meandering downstream with statistical certainty.

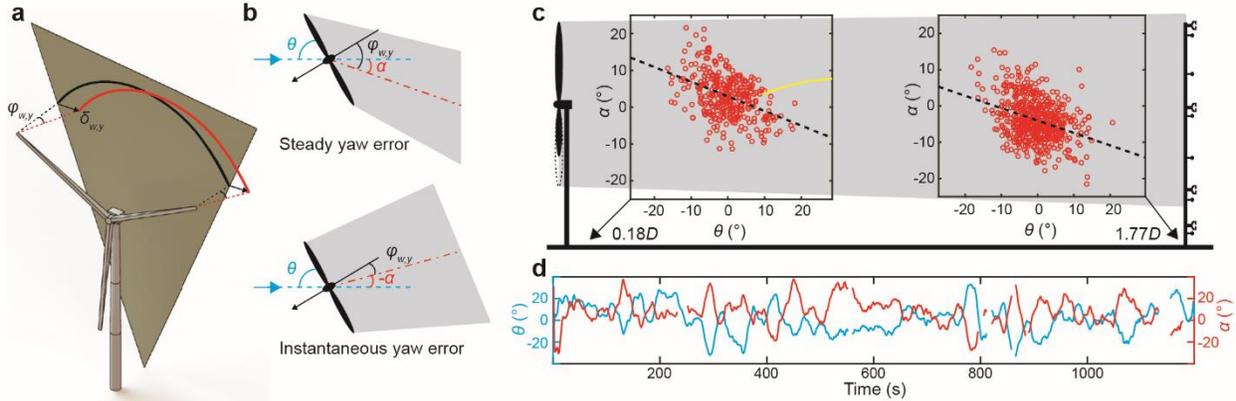

**Fig. 3 | Spanwise wake modulation. a**, Schematic showing how spanwise wake modulation is quantified. **b**, Diagrams of (top) average wake deflection under steady yaw error and (bottom) instantaneous deflection under instantaneous yaw error. **c**, Instantaneous spanwise wake deflection at two different locations downstream of the turbine: the light sheet position $0.18D$ downstream (left) and the met tower $1.77D$ downstream (right). Each red data point represents an average over 20 s of data, corresponding to the smoothing window applied to the wake envelope. The black dashed lines are least squares best fit lines and the yellow line is the trend for steady yaw wake deflection from Jiménez et al. 2010. **d**, Sample time series showing yaw error, $\theta$, and wake steer, $\alpha$, captured at the light sheet location.

Figure 4 shows the vertical wake modulation response to dynamic turbine operation. In Fig. 4a, vertical wake modulation is defined by $\delta_{w,y}$, displacement along the vertical direction, and $\varphi_{w,y}$, the angle of vertical deflection. It should be noted that, because only the top half of the wake is measured, the vertical modulation reported here is a combination of wake expansion and vertical deflection of the wake centerline. Figure 4b compares a sample time series of vertical wake modulation to blade pitch ($\beta$) and tip speed ratio ($\lambda$), showing a negative relationship with $\beta$ (correlation coefficient of -0.74 when the turbine is operating above region 2) and a positive relationship with $\lambda$ (correlation coefficient of 0.58). Combining these parameters in a contour plot, Fig. 4c shows nearly identical trends to those for the thrust coefficient shown in Fig. 2 of Gebraad et al. (2015) and reproduced in Supplementary Fig. 6. This similarity suggests that vertical wake modulation is largely contributed by wake expansion caused by changes in axial induction, which is a direct function of the thrust coefficient. This characteristic of vertical wake modulation enables the direct prediction of its behavior through blade pitch and tip speed ratio, which are readily available from SCADA data. Such prediction of vertical wake modulation can significantly improve the estimation of kinetic energy available for power generation, as thoroughly discussed in the section below.

Occasionally sudden decreases in vertical wake modulation are observed (Fig. 4d). These appear to occur when the wind speed is high (above approximately 13 m/s) and the turbine is operating in region 3. In this region, the control algorithm attempts to maintain the rotor speed at the rated level by adjusting the blade pitch. When the rotor speed increases past the rated value (these periods are indicated by gray bars in Fig. 4d), the control algorithm must slow it down, injecting energy back into the wake and causing an abrupt reduction in expansion. As our results show, this aspect of vertical wake modulation behavior is strongly dependent on the region of turbine operation and the control parameters recorded by the SCADA, enabling accurate prediction of this phenomenon and its effect on wake recovery.

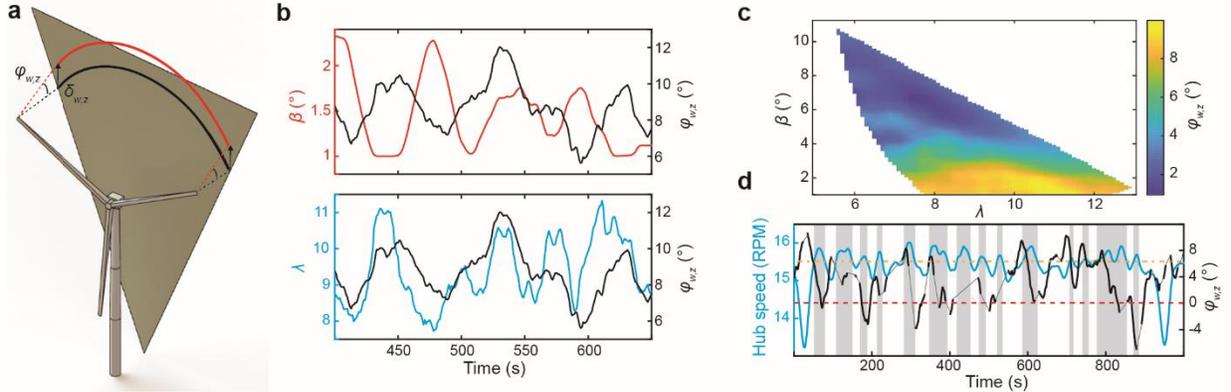

**Fig. 4 | Vertical wake modulation. a**, Schematic showing how vertical wake modulation is quantified. **b**, Sample time series showing vertical wake modulation with (top) blade pitch and (bottom) tip speed ratio. **c**, Interpolated and smoothed contour plot showing the relationship between vertical wake modulation, blade pitch, and tip speed ratio. **d**, Sample time series of hub speed and vertical wake modulation. Gray bars indicate periods where the hub speed is above the rated speed, 15.5 RPM, also indicated by a dash-dotted brown line. The red dashed line shows where the vertical wake modulation crosses zero. The thinner gray lines connecting the thick black lines represent vertical wake modulation data linearly interpolated through periods of intermittency, included to facilitate visualization.

**Influence on wake recovery**

To quantify the effect of dynamic wake modulation on the energy flux into the wake, a modified version of the streamtube method implemented by Lebron et al. (2012) is employed. Rather than a constant control volume, a moving and deforming control volume is used to capture the wake motion, where the boundary of the streamtube is the boundary of the wake, demarcated by the blade tip vortices. In a large wind farm, flux caused by the Reynolds shear stress is the main contributor to energy re-entrainment and wake recovery (Cal et al. 2010, Calaf et al. 2010, Hamilton et al. 2012, Lebron et al. 2012). To quantify the contribution of wake deformation to this term in the energy balance, the vertical and spanwise wake deflection velocities are calculated (Fig. 5a), and their fluctuating components are used along with the incoming and wake streamwise velocities from the SCADA and met tower, respectively, to calculate the corresponding flux terms, $\phi_y = \bar{u}_x \langle u_x' v_y' \rangle$ and $\phi_z = \bar{u}_x \langle u_x' v_z' \rangle$. These components of flux, normalized by the cube of the incoming wind speed, are plotted over the duration of the dataset in Fig. 5b. The contribution of each component varies with the region of operation. When the turbine is operating in region 3, the vertical energy flux induced by the vertical wake modulation is the main contribution to the total flux. In this region, as explained above, the regulation of the rotor speed causes abrupt changes in vertical wake modulation. In region 2, where less power is produced, the total flux caused by dynamic wake modulation is at its minimum due to the diminishing difference in velocity between the wake and the surrounding freestream flow. In region 2.5, the axial induction is at its maximum, as shown in Supplementary Fig. 7, and the vertical and spanwise components of modulation contribute nearly equally to the flux in this region.

In Fig. 5c the flux caused by dynamic wake modulation is compared to the total, vertical, and spanwise energy flux into the wake calculated using large eddy simulation by Cortina et al. (2016) for an isolated turbine in a neutrally stratified atmosphere. This plot shows that vertical flux caused by dynamic wake modulation can be up to 50% of the vertical energy flux calculated using the velocity components alone. In a large wind farm where a fully developed wind farm boundary layer has formed, the wake mixing that supplies energy to downwind turbines is primarily determined by vertical kinetic energy flux, as the energy in the horizontal direction has been depleted by the surrounding turbines (Cal et al. 2010, Calaf et al. 2010, Hamilton et al. 2012, Lebron et al. 2012, Peña & Rathmann 2014). Therefore, the enhanced mixing caused by vertical wake modulation can potentially lead to substantial increase of the wind power available in a large wind farm. In the spanwise direction, dynamic wake modulation can cause energy flux up to 10% of spanwise energy flux with no deformation included. Overall, dynamic wake modulation can contribute up to 20% more energy flux and an average of 11% more than that calculated without considering dynamic wake modulation. This additional energy flux has significant implications for wake mixing models, which currently do not account for dynamic wake modulation.

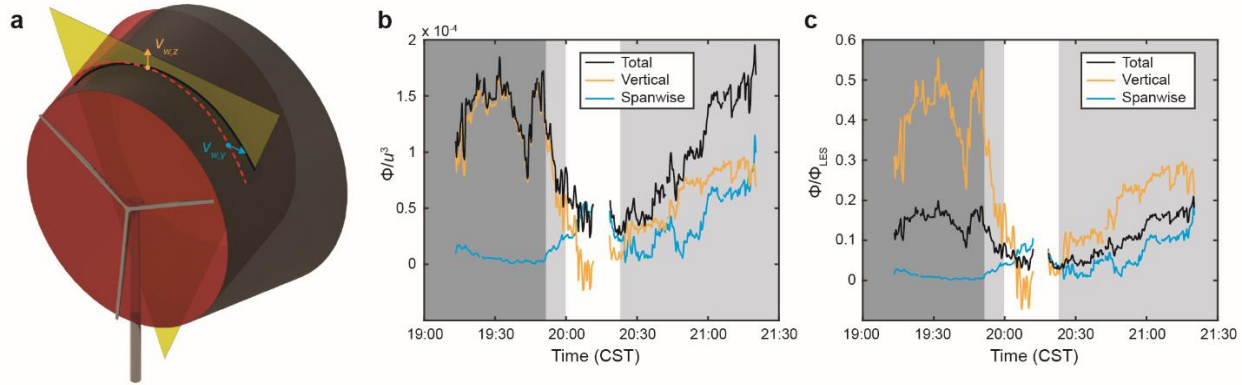

**Fig. 5 | Contribution of dynamic wake modulation to energy flux. a**, Schematic showing wake deflection velocity in the vertical and spanwise directions. The fluctuating components of these velocities are used to calculate the contribution of wake deformation to energy flux. **b**, Plot of the vertical, spanwise, and total wake modulation contributions to energy flux through a wake cross-section, normalized by the cube of the incoming velocity. The dark gray region indicates the time period where the turbine is operating in region 3, light gray in region 2.5, and white in region 2. **c**, Plot of dynamic wake modulation contributions to energy flux, normalized by the total flux per unit area in the vertical and spanwise directions reported by Cortina et al. (2016) for an isolated wind turbine.

**Discussion**

Through super-large-scale flow visualization using natural snowfall, we reveal and quantify dynamic wake modulation for the first time, which cannot be captured with any other existing techniques. This data is captured at unprecedented spatiotemporal resolution, and includes flow visualization over a 2-hour time period containing multiple regions of turbine operation and wind conditions. In the future, this high-resolution data can be used to validate numerical simulations of utility-scale wind turbine near wakes. Applying this visualization technique in the plane parallel to the rotor allows us to decompose the dynamic wake modulation into two modes: spanwise and vertical. Each mode contributes differently to the wake evolution. The spanwise mode is dependent on instantaneous yaw error and significantly affects the variability of the wake propagation direction, which is important for wind farm control schemes relying on wake steering. It should be noted that our results do not contradict the well-established behavior of wake steering as reported in many previous studies. Rather, these instantaneous deflections would be superimposed on top of any mean wake deflection, increasing the fluctuations in wake position. Although the current study only captures part of the blade tip vortex helix, preventing the differentiation between wake expansion and vertical centerline deflection in the vertical mode, future snow visualization measurements can be designed to capture a more complete section of the wake by expanding the light sheet and increasing the spotlight intensity. Nevertheless, the relationship between vertical wake modulation and tip speed ratio and blade pitch, which are related to turbine axial induction, suggests wake expansion provides the main contribution to the vertical mode. Both spanwise and vertical wake modulations influence wake recovery, crucial for accurate modelling of wind farm power output, by enhancing mixing at the wake boundary. Modulation in the vertical direction is particularly important for large wind farms where most energy is entrained from above the farm, so the ability to control this behavior would be highly beneficial for wind farm optimization.

Significantly, dynamic wake modulation is directly dependent on parameters that are already being measured by the SCADA system of the turbine, so it can easily be integrated into the control algorithms of existing wind farms. These relationships suggest that the parameters that determine dynamic wake modulation (e.g., blade pitch) can be modified to enhance wake mixing and provide more power to downwind turbines. For example, blade pitch is already used to control the turbine loading under rapidly changing wind conditions. Further, Gebraad et al. (2015) discuss the potential of using changes in blade pitch to regulate the thrust coefficient for the purposes of axial induction control without significantly reducing the power coefficient. Therefore blade pitch could also be used to control dynamic wake modulation with the goal of enhancing wake mixing and recovery to improve the overall energy extraction efficiency of the entire wind farm.

Additionally, dynamic wake modulation can be incorporated into layout design decisions for future wind farms. Meyers and Meneveau (2012) conducted an analytical optimization of wind turbine spacing and found that cost effectiveness is maximized when turbines are spaced $\sim 15D$ apart, which is significantly higher than the current standard for turbine spacing. However, their study was based on top-down models that account for the effect of wake turbulence on mixing in the wind turbine boundary layer (Calaf et al. 2010, Meneveau 2012), but do not include the additional mixing caused by dynamic wake modulation. If this additional mixing were included, the wind speed experienced by the wind farm would increase, decreasing the necessary spacing. The idea that enhanced mixing can decrease spacing is confirmed by the computational study conducted by Santhanagopalan et al. (2018) which found that increasing the incoming flow turbulence intensity decreases the optimal turbine spacing in a single turbine column. Additionally, Meyers and Meneveau (2012) found that when the turbines are operating in region 3, the optimal spacing decreases due to a decrease in thrust coefficient and a corresponding increase in boundary layer wind speed. This finding suggests that changing the wake behavior can allow closer turbine spacing. Controlling dynamic wake modulation could be one way to expedite wake recovery and reduce the optimal turbine spacing, enabling more efficient land area use.

In the current study, we validate our near-wake findings with data from the met tower at almost $2D$ downstream, but future work can be done to understand the propagation of observed features farther downstream. Observing these behaviors at a distance similar to turbine spacing in a wind farm ($6-8D$ downstream) would be beneficial for quantifying their effects on downwind turbines, and can be achieved through simultaneous measurements with other conventional field measurement techniques (e.g., lidar). Future work should also be done to integrate the findings of this study into numerical simulations. The omission of mixing caused by dynamic wake modulation from top-down models could be partially responsible for the underestimation of the power output of large wind farms such as the Horns Rev wind farm in Denmark (Stevens & Meneveau 2017). By incorporating dynamic wake modulation into wake models, wind farm power output can be predicted more accurately and optimized more effectively, helping wind energy become a more economical renewable energy option.

**Methods**

**Eolos field site.** The experiment was conducted at the University of Minnesota Eolos field site in Rosemount, MN (Supplementary Fig.1). The site consists of a heavily instrumented 2.5 MW Clipper Liberty C96 wind turbine and a 130 m met tower. The turbine is a three-bladed, horizontal-axis, pitch-regulated, variable speed machine with a 96 m rotor diameter mounted atop an 80 m tall support tower. A SCADA system is located at the hub, and strain gages are mounted around the tower base and along the blades. The SCADA system recorded incoming wind speed and direction at a frequency of 1 Hz and hub speed, blade pitch, power generated, and rotor direction at 20 Hz for this study. The met tower is located 170 m south of the turbine and comprises wind speed, direction, temperature, and humidity sensors at six elevations ranging from 7 m to 129 m (details provided in Supplementary Fig. 1). All the sensors record data 24 hours a day, which is stored on database servers.

**Flow visualization using natural snowfall.** Super-large-scale flow visualization using natural snowfall is described in detail in Hong et al. (2014), but a brief summary is provided here. Natural snowflakes serve as the environmentally benign seeding mechanism for a large volume in the near-wake of the turbine over several hours. They have strong light-scattering capabilities, and sufficient traceability for large-scale flow structures. Our previous publications have conducted detailed analysis of the traceability of snowflakes for large-scale flow measurements (Hong et al. 2014, Toloui et al. 2014, Nemes et al. 2017, Heisel et al. 2018 and Dasari et al. 2019). In the current study, snowflake patterns representing coherent flow structures are tracked rather than individual snowflakes. This specific concept is validated in Dasari et al. (2019).

**Experimental setup.** The flow visualization setup (Fig. 1a) is composed of an optical assembly for illumination and a camera. The optical assembly includes a 5-kW collimated searchlight with a 300 mm beam diameter (divergence < 0.3°) and a curved reflecting mirror to project the horizontal beam into a vertical light sheet. The sheet expansion angle is controlled by adjusting the mirror curvature. The local wind direction was used to align the light sheet perpendicular to the wind and parallel to the rotor with an angle of 96° clockwise from North, 17 m downstream of the turbine (Supplementary Fig. 2). As the wind direction changed throughout the experiment, the degree of misalignment between the light sheet and the rotor changed within the range of -14.5° to 16.2°. A Nikon D600 camera with a 50 mm f/1.2 Nikon lens was used to capture video data at a frame rate of 30 Hz and size of 1080 pixels × 1920 pixels. The camera was placed 166 m away from the light sheet and tilted 22.6° with respect to the ground in order to capture the entire vertical rotor span. The setup resulted in a field of view of 73 m × 129 m (spanwise × vertical).

**Experimental conditions.** The experiment took place between 19:00 and 22:00 CST on January 22$^{nd}$, 2018. The atmospheric and incoming wind conditions during this period were recorded using the met tower and SCADA system (Supplementary Fig. 3). The temperature at hub height stayed relatively constant between -3.8°C and -3.5°C. The variation in temperature between the bottom and top of the wake was approximately 0.8°C. The wind speed varied significantly over the course of the experiment, with instantaneous values between 2 m/s and 17 m/s at the hub, allowing the characterization of the wake under different turbine regions of operation. The wind direction also varied between approximately -30° and 20° clockwise from North. Due to the wind direction, the met tower was in the wake of the turbine throughout this time period.

**Void extraction from images.** Several image processing steps were applied to the raw images to extract the voids that indicate vortices in the flow, shown in Fig. 1b and in more detail focusing on the top tip vortices in Supplementary Fig. 4. First, the images were detilted to correct for distortion caused by the inclination angle of the camera using the process described in Dasari et al. (2019). A 500-frame moving average intensity image was calculated and subtracted from each image in the sequence. Histogram equalization was applied to enhance the contrast, and a 1.5 pixel median filter was applied to reduce noise (Supplementary Fig. 4b). The region outside of the light sheet was masked out. The images were then denoised using total variation denoising (Supplementary Fig. 4c). The images were binarized using a threshold calculated based on the average intensity of the image sequence (Supplementary Fig. 4d). The blades were removed from the non-thresholded images using a mask generated by averaging images that were highly correlated in the region of the blades. Median filter smoothing (Supplementary Fig. 4e) and Canny edge detection (Supplementary Fig. 4f) were also applied to the non-thresholded images. Binary regions that contained edges detected by the Canny filter were kept as snowflake voids caused by blade tip vortices in the wake, while ones that did not were removed as noise (Supplementary Figs. 4g and 4h). As the last step, the snow voids induced by blade tip vortices were extracted from the binary image based on their characteristic size, eccentricity, orientation, and location in the image. These parameters differentiated the tip vortex induced voids from the snow voids caused by other vortices in the near wake (e.g., vortices shed from the turbine hub) and in the ambient atmospheric flow passing through the wake.

**Wake modulation quantification.** The blade tip vortices extracted from the images define the wake envelope that was used to calculate the wake deflection. The topmost tip vortex edge was detected and outliers more than three scaled median absolute deviations from the median were removed. The envelope was interpolated to a grid and smoothed using a mean filter of 20 s. This smoothing period was determined by dividing the characteristic length scale, $D$, by the minimum wake velocity to remove any fluctuations that were too small to accurately reflect changes in the whole wake. Each cross-section of the envelope was compared to the wake cross-section extracted from the SolidWorks model of the experimental setup. In the model, the cross-section was obtained by approximating the wake as a uniform cylinder emanating from the rotor and intersecting the light sheet (Supplementary Fig. 5). The cross-section changed when nacelle direction changed, but no changes in atmospheric conditions or the resulting expansion and deflection were accounted for in the model. Because the model cross-section is a cylinder intersecting a plane, it is elliptical by definition. The equation of the ellipse was obtained and the shift that provided the best nonlinear least squares fit to the ellipse was applied to the experimental wake cross-section. The components of the shift of this cross-section define the vertical and spanwise wake shifts $\delta_{w,z}$ and $\delta_{w,y}$.

**Energy flux calculation.** The energy flux is calculated using the fluctuating velocities of the top and sides of the wake. The wake velocities are calculated by taking the temporal derivative of the wake deflection in the vertical and spanwise directions, $v_{w,z} = d\delta_{w,z}/dt$ and $v_{w,y} = d\delta_{w,y}/dt$. These velocities are multiplied by the fluctuating component of the streamwise freestream velocity obtained from the speed recorded by the SCADA system. Note that the wind speed recorded at the hub is an approximation of the freestream, as it is within the induction zone of the rotor. However, because the flow data from the SCADA is available to turbine controllers, this data is used for our analysis. The wind speed is decomposed into streamwise and spanwise components ($u_x$ and $u_y$, respectively) using the nacelle direction. A moving average streamwise velocity, $\bar{u}_x$, is calculated using a 30 minute window to account for changes in the atmospheric conditions. The fluctuating component is obtained by subtracting the moving average, $u'_x = u_x - \bar{u}_x$. The Reynolds shear stress caused by wake deformation is then defined as $\overline{u'_x v'}_{w,z}$ on the top surface and $\overline{u'_x v'}_{w,y}$ on the side surfaces. These Reynolds stresses contribute a flux per unit area of $\phi = \bar{u}_x \overline{u'_x v'}_w$ on each surface, where the $\bar{u}_x$ values representing freestream velocity are taken from the SCADA and the $\bar{u}_x$ values representing the wake velocity are taken from the met tower (which stays within the turbine wake throughout the experiment).

**Acknowledgements**

This work was supported by the National Science Foundation CAREER award (NSF-CBET-1454259), Xcel Energy through the Renewable Development Fund (grant RD4-13) as well as IonE of University of Minnesota. We also thank the students and the engineers from St Anthony Falls Laboratory, including S. Riley, T. Dasari, B. Li, Y. Wu, J. Tucker, C. Ellis, J. Marr, C. Milliren and D. Christopher for their assistance in the experiments.

Heisel, M., Dasari, T., Liu, Y., Hong, J., Coletti, F., & Guala, M. (2018). The spatial structure of the logarithmic region in very-high-Reynolds-number rough wall turbulent boundary layers. *Journal of Fluid Mechanics*, *857*, 704–747. https://doi.org/10.1017/jfm.2018.759

Hong, J., Toloui, M., Chamorro, L. P., Guala, M., Howard, K., Riley, S., Tucker, J., & Sotiropoulos, F. (2014). Natural snowfall reveals large-scale flow structures in the wake of a 2.5-MW wind turbine. *Nature Communications*, *5*. https://doi.org/10.1038/ncomms5216

Jiménez, Á., Crespo, A., & Migoya, E. (2010). Application of a LES technique to characterize the wake deflection of a wind turbine in yaw. *Wind Energy*, *13*, 559–572. https://doi.org/10.1002/we.380

Larsen, T. J., Madsen, H. A., Larsen, G. C., & Hansen, K. S. (2013). Validation of the dynamic wake meander model for loads and power production in the Egmond aan Zee wind farm. *Wind Energy*, *16*, 605–624. https://doi.org/10.1002/we

Larsen, G. C., Madsen, H. A., Thomsen, K., & Larsen, T. J. (2008). Wake meandering: A pragmatic approach. *Wind Energy*, *11*(4), 377–395. https://doi.org/10.1002/we.267

Lebron, J., Castillo, L., & Meneveau, C. (2012). Experimental study of the kinetic energy budget in a wind turbine streamtube. *Journal of Turbulence*, *13*, N43. https://doi.org/10.1080/14685248.2012.705005

Leishman, J. G. (2002). Challenges in modelling the unsteady aerodynamics of wind turbines. *Wind Energy*, *5*, 85–132. https://doi.org/10.1002/we.62

Lignarolo, L. E. M., Ragni, D., Scarano, F., Simão Ferreira, C. J., & Van Bussel, G. J. W. (2015). Tip-vortex instability and turbulent mixing in wind-turbine wakes. *Journal of Fluid Mechanics*, *781*, 467–493. https://doi.org/10.1017/jfm.2015.470

Lundquist, J. K., Duvivier, K. K., Kaffine, D., & Tomaszewski, J. M. (2019). Costs and consequences of wind turbine wake effects arising from uncoordinated wind energy development. *Nature Energy*, *4*(1), 26. https://doi.org/10.1038/s41560-018-0281-2

Machefaux, E., Larsen, G. C., Koblitz, T., Troldborg, N., Kelly, M. C., Chougule, A., … Rodrigo, J. S. (2016). An experimental and numerical study of the atmospheric stability impact on wind turbine wakes. *Wind Energy*, *19*, 1785–1805. https://doi.org/10.1002/we

Magnusson, M., & Smedman, A. (1994). Influence of atmospheric stability on wind turbine wakes. *Wind Engineering*, *18*(3), 139–152.

Marden, J. R., Ruben, S. D., & Pao, L. Y. (2013). A model-free approach to wind farm control using game theoretic methods. *IEEE Transactions on Control Systems Technology*, *21*(4), 1207–1214. https://doi.org/10.1109/TCST.2013.2257780

Meyers, J., & Meneveau, C. (2012). Optimal turbine spacing in fully developed wind farm. *Wind Energy*, *15*, 305–317. https://doi.org/10.1002/we

Meneveau, C. (2012). The top-down model of wind farm boundary layers and its applications. *Journal of Turbulence*, *13*, N7. https://doi.org/10.1080/14685248.2012.663092

Mirocha, J. D., Rajewski, D. A., Marjanovic, N., Lundquist, J. K., Kosovic, B., Draxl, C., … Churchfield, M. J. (2015). Investigating wind turbine impacts on near- wake flow using profiling lidar data and large-eddy simulations with an actuator disk model. *Journal of Renewable and Sustainable Energy*, *7*, 043243. https://doi.org/10.1063/1.4928873

Munters, W., & Meyers, J. (2018). Dynamic strategies for yaw and induction control of wind farms based on large-eddy simulation and optimization. *Energies*, *11*(1). https://doi.org/10.1111/j.0022-4367.2004.00085.x

Nemes, A., Dasari, T., Hong, J., Guala, M., & Coletti, F. (2017). Snowflakes in the atmospheric surface layer: observation of particle-turbulence dynamics. *Journal of Fluid Mechanics*, *814*, 592–613. https://doi.org/10.1017/jfm.2017.13

Peña, A., & Rathmann, O. (2014). Atmospheric stability-dependent infinite wind-farm models and the wake-decay coefficient. *Wind Energy*, *17*, 1269–1285. https://doi.org/10.1002/we

Santhanagopalan, V., Rotea, M. A., & Iungo, G. V. (2018). Performance optimization of a wind turbine column for different incoming wind turbulence. *Renewable Energy*, *116*, 232–243. https://doi.org/10.1016/j.renene.2017.05.046